\documentclass[conference]{IEEEtran}
\usepackage{amsmath, nccmath}
\usepackage{cases} 
\usepackage{amssymb}
\usepackage{cite}
\usepackage{url}
\usepackage{xcolor}
\usepackage{cite,graphicx}
\usepackage{subfigure}
\usepackage{citesort}
\usepackage{fancyhdr}
\usepackage{mdwmath}
\usepackage{mdwtab}
\usepackage{caption}
\usepackage{amsthm}
\usepackage{algorithmic}
\usepackage{algorithm}
\usepackage{threeparttable}
\usepackage{makecell}
\usepackage{booktabs}
\usepackage{pifont}

\newtheorem{theorem}{Theorem}

\newtheorem{lemma}{Lemma}

\newtheorem{corollary}{Corollary}

\makeatletter
\def\ScaleIfNeeded{%
\ifdim\Gin@nat@width>\linewidth \linewidth \else \Gin@nat@width
\fi } \makeatother

\begin{document}

\title{CRB minimization for  PASS   Assisted ISAC}
\author{
\IEEEauthorblockN{ Haochen~Li\IEEEauthorrefmark{1}, Ruikang~Zhong\IEEEauthorrefmark{2}, Jiayi~Lei\IEEEauthorrefmark{3}, Pan~Zhiwen\IEEEauthorrefmark{1}, and Yuanwei~Liu\IEEEauthorrefmark{4}} \IEEEauthorblockA{
\IEEEauthorrefmark{1} National Mobile Communications Research Laboratory, Southeast University, Nanjing, China\\
\IEEEauthorrefmark{2} Xi'an Jiaotong University, Xi'an, China\\
\IEEEauthorrefmark{3} Beijing University of Posts and Telecommunications, Beijing, China\\
\IEEEauthorrefmark{4} The University of Hong Kong, Hong Kong
 } }

\maketitle
\begin{abstract}
A multiple waveguide PASS assisted integrated sensing and communication (ISAC) system is proposed, where the  base station (BS) is equipped with transmitting pinching antennas (PAs) and receiving uniform linear array (ULA) antennas. The PASS-transmitting-ULA-receiving (PTUR) BS transmits the communication and sensing signals through the stretched PAs on waveguides and collects the echo sensing signals with the mounted ULA. Based on this configuration, a target sensing Cramér–Rao Bound (CRB) minimization problem is formulated under communication quality-of-service (QoS) constraints, power budget constraints, and PA deployment constraints. An alternating optimization (AO) method is employed to address the formulated non-convex optimization problem. Simulation results demonstrate that the proposed PASS assisted ISAC framework achieves superior performance over benchmark schemes.

\end{abstract}

\begin{IEEEkeywords}
Cramér–Rao Bound (CRB), integrated sensing and communications
(ISAC), pinching antenna systems (PASS).
\end{IEEEkeywords}\vspace{-0.1cm}

\section{Introduction}\label{I}
Integrated sensing and communication (ISAC) has emerged as a promising solution to enable efficient use of spectrum, power, and hardware resources in next-generation wireless networks. Adopting flexible antenna architectures in ISAC systems has been extensively investigated to fully exploit their potential benefits~\cite{10707252}. However, challenges such as line-of-sight~(LoS) path blockage and severe wireless propagation loss remain persistent obstacles. Although RIS-aided ISAC systems are capable of establishing alternative LoS paths between the base station (BS) and the sensing targets or communication users, the passive nature of RIS inherently leads to the notorious double fading effect, which significantly exacerbates the propagation loss~\cite{basar2021reconfigurable}. In contrast, MA or FA assisted ISAC systems are not subject to the double-fading issue due to their active nature. Nevertheless, they are still vulnerable to LoS path blockages between the BS and the sensing targets or communication users~\cite{wong2023fluid}.

The aforementioned challenges can be effectively addressed by the newly proposed flexible antenna architecture—pinching antenna systems (PASS)~\cite{yang2025pinching}. The core idea of PASS is to deploy dielectric waveguides across the service area, onto which dielectric particles—referred to as pinching antennas (PAs)—can be flexibly applied~\cite{liu2025pinching}. On one hand, the large deployment region and adjustable placement of PAs enable PASS to dynamically establish LoS paths to users and targets. On the other hand, by extending waveguides throughout the service area, PASS allows PAs to be deployed in close proximity to users and targets, thereby mitigating propagation loss that typically scales with the square of the transmission distance~\cite{suzuki2022pinching}. 

Several recent studies have explored PASS under different ISAC configurations and performance objectives~\cite{qin2025joint,khalili2025pinching,ouyang2025rate,mao2025multi}. the authors of~\cite{qin2025joint} formulated a joint PA placement and power allocation problem, aiming to maximize communication rate while satisfying sensing and energy constraints, and solved it with a maximum entropy reinforcement learning algorithm. In study~\cite{khalili2025pinching}, a novel PASS assisted ISAC architecture was proposed, where dynamically activated PAs achieve target diversity. By modeling radar cross section as a random variable and adopting outage probability as a reliability metric, the PA activation problem was solved using a successive convex approximation (SCA) method. The authors of~\cite{ouyang2025rate} analyzed the fundamental limits of PASS assisted ISAC systems, deriving closed form expressions for achievable communication and sensing rates under various beamforming objectives, including sensing centric, communication centric, and Pareto optimal designs. The authors of~\cite{mao2025multi} studied a multi-waveguide PASS assisted ISAC framework, where transmit and receive PAs were jointly optimized under rate, SNR, power, and placement constraints. A fine tuning approximation and SCA based method were proposed, demonstrating the feasibility of large scale PASS deployment in practical ISAC systems. 

Existing PASS assisted ISAC studies~\cite{qin2025joint,khalili2025pinching,ouyang2025rate,mao2025multi} adopt \textit{sensing SNR} or \textit{sensing power} as performance metrics, owing to their ease of computation and direct relation to system parameters such as transmit power, beamforming gain, and antenna configuration. These metrics are well suited for early-stage system design, link budget analysis, and power allocation. Nevertheless, sensing SNR does not explicitly capture the accuracy of parameter estimation. In contrast, the CRB establishes a theoretical lower bound on the variance of unbiased estimators. To the best of our knowledge, no prior work has studied the CRB for target sensing in PASS assisted ISAC systems. To fill this gap, this paper adopts the CRB as the sensing metric for a PASS assisted ISAC system that simultaneously serves multiple communication users and senses a target. The main contributions of this work can be summarized as follows:
\begin{itemize}
        \item A general multiple waveguide PASS assisted ISAC system with the PASS-transmitting-ULA-receiving (PTUR) BS is proposed. A CRB minimization problem is formulated and then decomposed into a digital beamforming sub-problem and a pinching beamforming sub-problem with the AO method.
        \item For the digital beamforming sub-problem, the semidefinite relaxation (SDR) based optimization framework is developed to solve the non-convex sensing CRB minimization problem with rank-one constraint.
        \item For the pinching beamforming sub-problem, an AO framework is developed, combining penalty, SCA, and element-wise optimization techniques to address the non-convex and tightly coupled optimization of PA locations.
\end{itemize}\vspace{-0.1cm}

\section{System Model}\label{II}
\begin{figure} [htbp]
\centering\vspace{-0.3cm}
\includegraphics[width=0.33\textwidth]{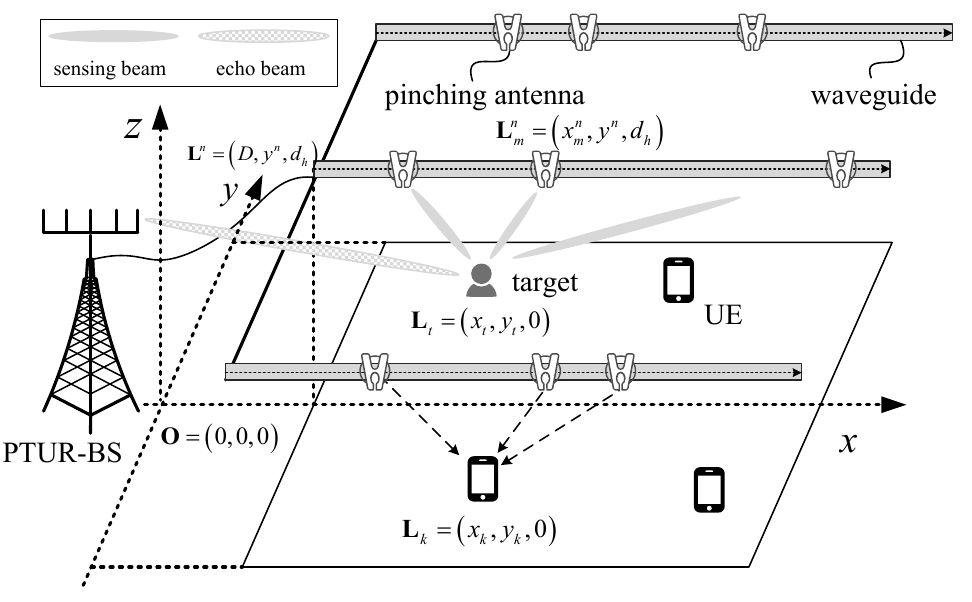}
 \caption{The proposed PASS assisted ISAC system.
  }\vspace{-0.3cm}
 \label{system_model}
\end{figure}
As shown in Fig.~\ref{system_model}, we consider a  PASS assisted  ISAC system, where the BS adopts the PTUR architecture, i.e., the signals are transmitted with a pinching antenna and the sensing echo is received with a ULA.  The dual function BS simultaneously communicates with $K$ user  and probes a target in the service area denoted by $\mathcal{A}$ with a $D_x \times D_y$ rectangle shape. The the location of the $k$-th UE and the target are $\boldsymbol{l}_k=\left[x_k,y_k,0\right]^\mathrm{T}, \forall k\in \mathcal{K},$ and $\boldsymbol{l}_t=\left[x_t,y_t,0\right]^\mathrm{T}$,respectively.  The PASS is equipped with $N$ waveguides and $M$ pinching antennas are deployed on each waveguide. All waveguides are elevated at height $d_h$ while extending along the $x$-aixs, and the fed point of the $n$-th waveguide is $\boldsymbol{l}^n=\left[D,y^n,d_h\right]^\mathrm{T}$. The location of the $m$-th pinching antenna on the $n$-th waveguide is $\boldsymbol{l}_m^n=\left[x_m^n,y^n,d_h\right]^\mathrm{T}$. In PASS, the position of each PA is subject to the following constraints\vspace{-0.1cm}
\begin{equation}\label{X}
\begin{aligned}
\mathcal{X}=\left\{ {\begin{array}{*{20}{c}}
{x_m^n,}&\vline& {D \le x_m^n \le D+L}\\[-0.05cm]
{ \forall n, m}&\vline& {x_m^n - x_{m-1}^n \ge \delta, \ \text{if} \ m>1}
\end{array}} \right\},
\end{aligned}\vspace{-0.1cm}
\end{equation}
where $L$ denotes the length of the waveguide. $\delta$ is the minimum gap between PAs.\vspace{-0.15cm}

\subsection{PASS Channel Model} 
The channels of PASS systems contain both the  wireless PA-UE channels and the in-waveguide channels. The wireless channel between the PAs on the $n$-th waveguide and user $k$ can be expressed as \vspace{-0.1cm}
\begin{equation}
\mathbf{h}^n_k=\left(2\kappa_c\right)^{-1}\!\Big[\frac{e^{-j \kappa_c r_{k,1}^n}} {r_{k,1}^n},\frac{e^{-j \kappa_c r_{k,2}^n}} {r_{k,2}^n},\cdots,\frac{e^{-j \kappa_c r_{k,M}^n}} {r_{k,M}^n}\Big]^\mathrm{T}\!,\vspace{-0.1cm}
\end{equation}
where $\kappa_c={2\pi f_c}/{c}={2\pi}/{\lambda_c}$ is the wavenumber. $r_{k,m}^n=\left\|\boldsymbol{l}_m^n-\boldsymbol{l}_k\right\|_2$ stands for the distance between the $m$-th PA on the $n$-th waveguide and user $k$. Then, the overall wireless channel between the PAs and user $k$ can be expressed as
\begin{equation}
\hat{\mathbf{h}}_k=\Big[\left(\mathbf{h}^1_k\right)^\mathrm{T},\left(\mathbf{h}^2_k\right)^\mathrm{T},\cdots,\left(\mathbf{h}^N_k\right)^\mathrm{T}\Big]^\mathrm{T}\in\mathbb{C}^{MN\times 1}.
\end{equation}
Similarly, the wireless channel between the PAs on the $n$-th waveguide and target can be given as \vspace{-0.1cm}
\begin{equation}
\mathbf{h}^n_t=\left(2\kappa_c\right)^{-1}\!\Big[\frac{e^{-j \kappa_c r_{t,1}^n}} {r_{t,1}^n},\frac{e^{-j \kappa_c r_{t,2}^n}} {r_{t,2}^n},\cdots,\frac{e^{-j \kappa_c r_{t,M}^n}} {r_{t,M}^n}\Big]^\mathrm{T}\!,\vspace{-0.1cm}
\end{equation}
where $r_{t,m}^n=\left\|\boldsymbol{l}_m^n-\boldsymbol{l}_t\right\|_2$ stands for the distance between the $m$-th PA on the $n$-th waveguide and target. Then, the wireless channel between the PAs and the target can be expressed as
\begin{equation}
\hat{\mathbf{h}}_t=\Big[\left(\mathbf{h}^1_t\right)^\mathrm{T},\left(\mathbf{h}^2_t\right)^\mathrm{T},\cdots,\left(\mathbf{h}^N_t\right)^\mathrm{T}\Big]^\mathrm{T}\in\mathbb{C}^{MN\times 1}.
\end{equation}
The in-waveguide channel is in the form of a block-diagonal matrix, which can be expressed as \vspace{-0.1cm}
\begin{equation}
\begin{aligned}
\mathbf{F}&=\text{Bdiag}\left(\mathbf{f}_1,\mathbf{f}_2,\cdots,\mathbf{f}_N\right)^\mathrm{T}\in\mathbb{C}^{N\times MN},
\end{aligned}\vspace{-0.1cm}
\end{equation}
where $\mathbf{f}_n=\left[f_1^n,f_2^n,\cdots,f_M^n\right]^\mathrm{T}\in\mathbb{C}^{M\times 1}$ with $f_m^n$ stands for the phase difference between the signal transmitted from the $m$-th PA of the $n$-th waveguide and the in-put signal for the $n$-th waveguide. Specifically, it can be expressed as\vspace{-0.2cm} 
\begin{equation}
f_m^n = \frac{1}{\sqrt{M}}e^{-j\kappa_gd_m^n},\vspace{-0.2cm}
\end{equation}
where $\kappa_g={2\pi}/{\lambda_g}$ is the guided wavenumber, and $\lambda_g=\lambda_c/n_{e}$ is the guided wavelength of the waveguide with effective refractive index $n_{e}$. $d_{m}^n=\left\|\boldsymbol{l}^n-\boldsymbol{l}_m^n\right\|_2$ stands for the distance between the $m$-th PA on the $n$-th waveguide and  the fed point of the $n$-th waveguide. Then, the overall channel between  PASS and user $k$ can be given as \vspace{-0.2cm}
\begin{equation}\label{h_k}
\mathbf{h}_k=\mathbf{F}\hat{\mathbf{h}}_k\in\mathbb{C}^{N\times 1}.\vspace{-0.1cm}
\end{equation}
The overall channel between the PASS and target is \vspace{-0.1cm}
\begin{equation}
\mathbf{h}_t=\mathbf{F}\hat{\mathbf{h}}_t\in\mathbb{C}^{N\times 1}.\vspace{-0.2cm}
\end{equation}

\subsection{Signal Model} 
During downlink ISAC, the PAs send communication and sensing signals to simultaneously serve all communication UEs and probe the target. The echo signals reflected by the target is collect by the ULA of the BS. At time slot $t$, the PA transmission signal can be expressed as 
\begin{equation}
\mathbf{s}\left(t\right)=\mathbf{W}\mathbf{c}\left(t\right)+\mathbf{c}_s\left(t\right),
\end{equation}
where $\mathbf{W}=\left[\mathbf{w}_1,\mathbf{w}_2,\cdots,\mathbf{w}_K\right]\in\mathbb{C}^{N\times K}$ is the beamforming matrix for communication users, with $\mathbf{w}_k\in\mathbb{C}^{N\times 1}$ denoting the beamformer assigned to user $k$, $\forall k \in \mathcal{K}$. $\mathbf{c}\left(t\right)=\left[{c}_1\left(t\right),{c}_2\left(t\right),\cdots,{c}_K\left(t\right)\right]^\mathrm{T}\in\mathbb{C}^{K\times 1}$ is the communication signal transmitted at time slot $t$, with $c_k\left(t\right)\in\mathbb{C}$ denoting the signal for user $k$, $\forall k \in \mathcal{K}$. The communication signals  for different users are independent, thus the normalized communication signal satisfies $\mathbb{E}\left\{\mathbf{c}\left(t\right)\mathbf{c}\left(t\right)^\mathrm{H}\right\}=\mathbf{I}_K$. $\mathbf{c}_s$ is the dedicate sensing signal with covariance matrix $\mathbf{R}_s$. The covariance matrix of the PA transmission signal is \vspace{-0.1cm}
\begin{equation}
        \mathbf{R}=\mathbb{E}\big\{\mathbf{s}\left(t\right)\mathbf{s}\left(t\right)^\mathrm{H}\big\}=\mathbf{W}\mathbf{W}^\mathrm{H}+\mathbf{R}_s.\vspace{-0.2cm}
\end{equation}

\subsection{Communication Model and Sensing Model}
At time slot $t$, the received signal of user $k$ is 
\begin{equation}
\begin{aligned}
        {y}_k\left(t\right)=\sum\nolimits_{i}\mathbf{h}_k^\mathrm{H}\mathbf{w}_i{c}_i\left(t\right)+\mathbf{h}_k^\mathrm{H}\mathbf{c}_s\left(t\right)+n_k\left(t\right),
\end{aligned}
\end{equation}
where ${n}_k\left(t\right) \sim \mathcal{CN}({0}, \sigma_0^2)$ denotes the additive white Gaussian noise at the receiver of user $k$. Treating the received sensing signal as interference, the communication SINR is\vspace{-0.1cm}
\begin{equation}\label{CommunicationSINR}
        \text{SINR}_{k} = \frac{|\mathbf{h}_{k}^\mathrm{H}\mathbf{w}_{k}|^2}{\sum\nolimits_{i\neq k}|\mathbf{h}_{k}^\mathrm{H}\mathbf{w}_{i}|^2+\mathbf{h}_{k}^\mathrm{H}\mathbf{R}_{s}\mathbf{h}_{k}+\sigma_0^2}.\vspace{-0.1cm}
\end{equation}

To carry out the sensing function, the BS sends probing signals with PAs and collects the reflected signals with the ULA. The echo signal received by the ULA at time slot $t$ is \vspace{-0.1cm}
\begin{equation}      
        \mathbf{y}_s\left(t\right)=\mathbf{G}\mathbf{s}\left(t\right)+\mathbf{H}_\mathrm{I}\mathbf{s}\left(t\right)+{\mathbf{n}}_s\left(t\right),\vspace{-0.1cm}
\end{equation}
where ${\mathbf{n}}_s\left(t\right)\sim \mathcal{CN}(\mathbf{0}, {\sigma}_s^2\mathbf{I}_{M_r})$ denotes the additive white Gaussian noise vector at the receiving ULA. $\mathbf{H}_\mathrm{I}\in\mathbb{C}^{M_r\times MN}$ stands for the self-interference channel between the PAs and the receiving ULA. $\mathbf{G}$ is the PASS-target-ULA two-hop sensing channel. It can be  expressed as \vspace{-0.1cm}
\begin{equation}
        \mathbf{G}=\beta\mathbf{a}\left(\theta\right)\mathbf{h}_t^\mathrm{H},\vspace{-0.1cm}
\end{equation}
where $\beta$ incorporates the target reflection factor and  the path loss between the BS ULA and target. $\mathbf{a}\left(\theta\right)$ is the steering vector of the $M_r$-element ULA with half wavelength element distance, i.e.,\vspace{-0.1cm}
\begin{equation}
    \mathbf{a}\left(\theta\right)=[1,e^{-jk_c \sin\theta},\cdots,e^{-jk_c\left(M_r-1\right)\sin\theta}]^\mathrm{T},   \vspace{-0.1cm} 
\end{equation}
where $\theta$ represents the angle of the target with respect to the receiving ULA. 

Since the PAs in PASS are placed farther away from the BS receiving antennas, the impact of self-interference is significantly reduced in the proposed PASS assisted ISAC system. Therefore, self-interference is neglected in this work. The echo signal received by the BS ULA at time slot $t$ is 
\begin{equation}      
        \mathbf{y}_s\left(t\right)=\mathbf{G}\mathbf{s}\left(t\right)+\mathbf{n}_s\left(t\right).
\end{equation}
The BS collects the echo signals over $T$ time slot\vspace{-0.1cm}
\begin{equation}
        \mathbf{Y}_s=\mathbf{G}\mathbf{S}+\mathbf{N}_s,\vspace{-0.1cm}
\end{equation}
where $\mathbf{Y}_s=[ \mathbf{y}_s\left(1\right),\mathbf{y}_s\left(2\right), \cdots , \mathbf{y}_s\left(T\right)]$, $\mathbf{S}=[ \mathbf{s}\left(1\right),\mathbf{s}\left(2\right), \cdots , \mathbf{s}\left(T\right)]$, and $\mathbf{N}_s=[ \mathbf{n}_s\left(1\right),\mathbf{n}_s\left(2\right), \cdots , \mathbf{n}_s\left(T\right)]$. Define the parameter vector as \vspace{-0.1cm}
\begin{equation}
        \boldsymbol{\eta} = \left[x_t,y_t,\Re\left(\beta\right),\Im\left(\beta\right)\right]^\mathrm{T}\in\mathbb{R}^{4\times 1}.\vspace{-0.1cm}
\end{equation} 
The FIM of $\boldsymbol{\eta}$ is \vspace{-0.1cm}
\begin{equation}\label{eqn:FIM}
        \mathbf{J}=\begin{bmatrix}
         \mathbf{J}_{11} & \mathbf{J}_{12};\quad
         \mathbf{J}_{12}^\mathrm{T} & \mathbf{J}_{22} 
    \end{bmatrix}\in\mathbb{R}^{4\times 4},\vspace{-0.1cm}
\end{equation}
The detailed expressions of ${\bf{J}}_{11}$, ${\bf{J}}_{12}$, and ${\bf{J}}_{22}$ are derived in Appendix~A. The CRB of estimating target location $x_t$ and  $y_t$ can be given as \vspace{-0.1cm}
\begin{equation}\label{eqn:CRB}
    \text{CRB}\left(x_t, y_t\right) = \left({\bf{J}}_{11}-{\bf{J}}_{12}{\bf{J}}_{22}^{-1}{\bf{J}}_{12}^T\right)^{-1}\in \mathbb{R}^{2 \times 2}.\vspace{-0.2cm}
\end{equation}
\subsection{Problem Formulation}
The objective is to minimize the trace of the CRB matrix, which characterizes the lower bound on the localization mean squared error (MSE) of the target position $(x_t, y_t)$. The corresponding optimization problem is formulated as follows\vspace{-0.1cm}
\begin{subequations}\label{problem:CRB}
    \begin{align}        
        \min_{\mathbf{X},\mathbf{W}, \mathbf{R}_s} \quad &  \text{tr}\left(\text{CRB}\right) \\[-0.05cm]
        \label{constraint:PASS}
        \mathrm{s.t.} \quad & \left[\mathbf{X}\right]_{mn}=x_m^n \in \mathcal{X}, \forall m \in \mathcal{M},n\in \mathcal{N},  \\[-0.05cm]
        \label{constraint:SINR}
        & \text{SINR}_{k} \ge \gamma_{k}, \forall k \in \mathcal{K},\\[-0.05cm]
        \label{constraint:power}
        & \text{tr}\left(\mathbf{W}\mathbf{W}^H+\mathbf{R}_{s}\right) \le P, \mathbf{R}_{s}  \succeq  0,
    \end{align}\vspace{-0.6cm}
\end{subequations}

\noindent where $\mathbf{X}$ is the PA deployment variable matrix, whose entry on the $m$-th row and $n$-th column is $x_m^n$. Constraint~\eqref{constraint:PASS} ensures that the placement of PAs complies with the hardware and electromagnetic limits defined in~\eqref{X}.
Constraint~\eqref{constraint:SINR} guarantees the communication  QoS  requirements for all users in the set $\mathcal{K}$.
Constraint~\eqref{constraint:power} imposes a total transmit power budget $P$ for the PASS system.\vspace{-0.15cm}

\section{Proposed Alternating Optimization Algorithm}\label{III}

To efficiently solve the non-convex problem in \eqref{problem:CRB}, we propose an AO algorithm to sequentially optimize $\{\mathbf{W},\mathbf{R}_s\}$ and $\mathbf{X}$.\vspace{-0.15cm}
\subsection{Optimize $\mathbf{W}$  and $\mathbf{R}_s$ with Fixed $\mathbf{X}$}\label{subsection_III_A}

When the PA positions $\mathbf{X}$ are fixed, the optimization problem becomes\vspace{-0.1cm}
\begin{subequations}\label{problem:CRB_WR}
    \begin{align}        
        \min_{ \mathbf{W}, \mathbf{R}_s} \quad &  \text{tr}\left(\text{CRB}\left(x_t, y_t\right)\right) \\[-0.05cm]
        \mathrm{s.t.} \quad & \eqref{constraint:SINR},\eqref{constraint:power}, 
    \end{align}\vspace{-0.6cm}
\end{subequations}

\noindent To further simplify the  sub-problem, introduce an auxiliary semi-definite matrix $\mathbf{U}$ which follows\vspace{-0.1cm}
\begin{equation}
        \text{tr}(\mathbf{U}^{-1})\ge\text{tr}\left(\text{CRB}\right)=\text{tr}(({\bf{J}}_{11}-{\bf{J}}_{12}{\bf{J}}_{22}^{-1}{\bf{J}}_{12}^T)^{-1}).\vspace{-0.1cm}
\end{equation}
Note that the function $\text{tr}(\mathbf{M}^{-1})$ is monotonic over any semi-definite matrix $\mathbf{M}$, thus we have $\left({\bf{J}}_{11}-\mathbf{U}\right)-{\bf{J}}_{12}{\bf{J}}_{22}^{-1}{\bf{J}}_{12}^T\succeq 0$, which can be expressed as following linear matrix inequality\vspace{-0.1cm}
\begin{equation}\label{eqn:LIM}
        \begin{bmatrix}
         \mathbf{J}_{11}-\mathbf{U} & \mathbf{J}_{12}; \quad
         \mathbf{J}_{12}^\mathrm{T} & \mathbf{J}_{22} 
    \end{bmatrix}\succeq 0.\vspace{-0.1cm}
\end{equation}
The problem~\eqref{problem:CRB_WR} can be reformulated as\vspace{-0.1cm}
\begin{subequations}\label{problem:CRB_WR_U}
    \begin{align}        
        \min_{\mathbf{U}, \mathbf{W}, \mathbf{R}_s} \quad &  \text{tr}\left(\mathbf{U}^{-1}\right) \\[-0.05cm]
        \mathrm{s.t.} \quad &~\eqref{constraint:SINR},~\eqref{constraint:power},~\eqref{eqn:LIM},\\[-0.05cm] 
        & \mathbf{R}_{s}, \mathbf{U}  \succeq  0.
    \end{align}\vspace{-0.6cm}
\end{subequations}

\noindent The reformulated problem is still  non-convex due to the  non-convex constraint~\eqref{constraint:SINR}. This constraint can be rewritten as \vspace{-0.1cm}
\begin{equation}\label{RW_SINR}
        \frac{1}{\gamma_{k}}|{\mathbf{h}}_k^\mathrm{H}{\mathbf{w}}_{k}|^2 \ge \sum\nolimits_{i\ne k}|{\mathbf{h}}_k^\mathrm{H}{\mathbf{w}}_{i}|^2+{\mathbf{h}}_k^\mathrm{H}{\mathbf{R}}_{s}{\mathbf{h}}_k+\sigma_0^2. \vspace{-0.1cm}
\end{equation}
Define ${\mathbf{W}}_k={\mathbf{w}}_k{\mathbf{w}}_k^\mathrm{H}$, $\forall k \in \mathcal{K}$. Constraint~\eqref{constraint:SINR} can be further rewritten as \vspace{-0.1cm}
\begin{equation}\label{RW_SINR_matrix}
\begin{aligned}
     \frac{1+\gamma_{k}}{\gamma_{k}}&\text{tr}\left({\mathbf{h}}_k{\mathbf{h}}_k^\mathrm{H}{\mathbf{W}}_{k}\right)\ge\sum\nolimits_{i}\text{tr}\left({\mathbf{h}}_k{\mathbf{h}}_k^\mathrm{H}{\mathbf{W}}_{i}\right) \\[-0.05cm] &+\text{tr}\left({\mathbf{h}}_k{\mathbf{h}}_k^\mathrm{H}{\mathbf{R}}_{s}\right)+\sigma_0^2=\text{tr}\left({\mathbf{h}}_k{\mathbf{h}}_k^\mathrm{H}{\mathbf{R}}\right)+\sigma_0^2.   
\end{aligned}\vspace{-0.1cm}
\end{equation}
The constraint~\eqref{constraint:power} can be reformulated as\vspace{-0.1cm}
\begin{equation}\label{constraint:power_matrix} 
        \text{tr}\Big(\sum\nolimits_{k=1}^K\mathbf{W}_k+\mathbf{R}_{s}\Big) \le P\vspace{-0.1cm}
\end{equation}
Substitute~\eqref{RW_SINR_matrix},~\eqref{constraint:power_matrix}, and auxiliary matrices ${\mathbf{W}}_k, \forall k $ into problem~\eqref{problem:CRB_WR_U}, the  reformulated problem can be given as\vspace{-0.1cm}
\begin{subequations}\label{problem:CRB_WR_U_matrix}
    \begin{align}        
        \min_{\mathbf{U}, \left\{\mathbf{W}_k\right\}_{k=1}^K, \mathbf{R}_s} \quad &  \text{tr}\left(\mathbf{U}^{-1}\right) \\[-0.05cm]
        \mathrm{s.t.} \quad &~\eqref{eqn:LIM},~\eqref{RW_SINR_matrix},~\eqref{constraint:power_matrix}, \\[-0.05cm]\label{constraint:semidefinite_all}
        & \mathbf{R}_{s}, \mathbf{U}, \mathbf{W}_k, \forall k,  \succeq  0,\\[-0.05cm]
        \label{constraint:rank}
        & \text{rank}\left(\mathbf{W}_k\right)=1, \forall k.
    \end{align}\vspace{-0.6cm}
\end{subequations}

\noindent This optimization problem is non-convex due to the non-convex rank-one constraint~\eqref{constraint:rank}. By relaxing this constraint, the SDR of problem~\eqref{problem:CRB_WR_U_matrix} is\vspace{-0.1cm}
\begin{subequations}\label{problem:CRB_WR_U_matrix_SDR}
    \begin{align}        
        \min_{\mathbf{U}, \left\{\mathbf{W}_k\right\}_{k=1}^K, \mathbf{R}_s} \quad &  \text{tr}\left(\mathbf{U}^{-1}\right) \\[-0.1cm]
        \mathrm{s.t.} \quad &~\eqref{eqn:LIM},~\eqref{RW_SINR_matrix},~\eqref{constraint:power_matrix},~\eqref{constraint:semidefinite_all}.
    \end{align}\vspace{-0.6cm}
\end{subequations}

\noindent This optimization problem is convex and can be efficiently solved using CVX.  Suppose $\{\{\hat{\mathbf{W}}_k\}_{k=1}^K, \hat{\mathbf{R}}_s\}$ is the solution to problem~\eqref{problem:CRB_WR_U_matrix_SDR}. $\{\hat{\mathbf{W}}_k\}_{k=1}^K$ obtained from the SDR generally has a  rank that is higher than one. It has been proved that a set of solution $\{\{\bar{\mathbf{W}}_k\}_{k=1}^K, \bar{\mathbf{R}}_s\}$ that satisfies constraint~\eqref{constraint:rank} always can be constructed from the optimal solution of problem~\eqref{problem:CRB_WR_U_matrix_SDR}, i.e., $\{\{\hat{\mathbf{W}}_k\}_{k=1}^K, \hat{\mathbf{R}}_s\}$. Specifically, for user $k$, let $\hat{\mathbf{w}}_k$\vspace{-0.1cm}
\begin{equation}
        \bar{\mathbf{w}}_k=\big({\mathbf{h}}_k^\mathrm{H}\hat{\mathbf{W}}_k{\mathbf{h}}_k\big)^{-1}\hat{\mathbf{W}}_k{\mathbf{h}}_k.\vspace{-0.1cm}
\end{equation}
The constructed rank-one solution is $\bar{\mathbf{W}}_k=\bar{\mathbf{w}}_k\bar{\mathbf{w}}_k^\mathrm{H}$. Then the sensing covariance matrix can be calculated as \vspace{-0.1cm}
\begin{equation}
     \bar{\mathbf{R}}_s= \hat{\mathbf{R}}_s- \sum\nolimits_{i}\bar{\mathbf{W}}_i .\vspace{-0.1cm}
\end{equation}
It can be verified that the constructed $\{\{\bar{\mathbf{W}}_k\}_{k=1}^K, \bar{\mathbf{R}}_s\}$ can meet constraints~\eqref{RW_SINR_matrix},~\eqref{constraint:power_matrix},~\eqref{constraint:semidefinite_all}, and~\eqref{constraint:rank}, while  achieving the same objective value as solution $\{\{\hat{\mathbf{W}}_k\}_{k=1}^K, \hat{\mathbf{R}}_s\}$. Thus, $\{\{\bar{\mathbf{W}}_k\}_{k=1}^K, \bar{\mathbf{R}}_s\}$ is the optimal solution of problem~\eqref{problem:CRB_WR_U_matrix}.\vspace{-0.1cm}

\subsection{Optimize $\mathbf{X}$ with Fixed $\mathbf{W}$ and $\mathbf{R}_s$}\label{subsection_III_B}

Fixing $\mathbf{W}$ and $\mathbf{R}_s$, the pinching beamforming sub-problem reduces to optimizing the deployment $\mathbf{X}$ of the PASS antennas \vspace{-0.3cm}
\begin{subequations}\label{problem:CRB_X}
    \begin{align}        
        \min_{ \mathbf{X}} \quad &  \text{tr}(\text{CRB}) \\[-0.05cm]
        \mathrm{s.t.} \quad & \eqref{constraint:PASS}, \eqref{constraint:SINR}. 
    \end{align}\vspace{-0.6cm}
\end{subequations}

\noindent The PA deployment problem~\eqref{problem:CRB_X} is non-convex due to the complex relationship between the sensing CRB/communication SINR and the PA deployment. In the following, we introduce auxiliary matrices to decompose the optimization variables and adopt the AO technique along with the penalty method and SCA to tackle this problem. Note that the overall
channel between PASS and user $k$ given in~\eqref{h_k} can be expressed as \vspace{-0.1cm}
\begin{equation}
\begin{aligned}
    \mathbf{h}_k\!\!=\!\!\sum_{m=1}^M\!\!\left[\left[\mathbf{f}_1\right]_m\left[\mathbf{h}_k^1\right]_m,\cdots,\left[\mathbf{f}_N\right]_m\left[\mathbf{h}_k^N\right]_m\right]^\mathrm{T}\!\!=\!\!\sum_{m=1}^M\!\!\mathbf{h}_{k,m},
\end{aligned}   \vspace{-0.1cm}
\end{equation}
where $\mathbf{h}_{k,m}$ stands for the channel between the $m$-th PA on all $N$ waveguides and user $k$. Based on this, the channel between the $m$-th PA on all $N$ waveguides and all $k$ communication users can be expressed as\vspace{-0.1cm}
\begin{equation}
        \mathbf{H}_m=\left[\mathbf{h}_{1,m},\mathbf{h}_{2,m},\cdots,\mathbf{h}_{K,m}\right]\in \mathbb{C}^{N \times K}.\vspace{-0.1cm}
\end{equation}
Then, the channel between the PASS and all $k$ communication users can be expressed as\vspace{-0.1cm}
\begin{equation}
        \mathbf{H}=\left[\mathbf{h}_{1},\mathbf{h}_{2},\cdots,\mathbf{h}_{K}\right]=\sum\nolimits_{m=1}^M\mathbf{H}_m\in \mathbb{C}^{N \times K}.\vspace{-0.1cm}
\end{equation}
Define auxiliary matrices and vectors\vspace{-0.1cm} 
\begin{equation}
        \mathbf{Q}=\left[\mathbf{q}_{1},\mathbf{q}_{2},\cdots,\mathbf{q}_{K}\right]=\sum\nolimits_{m=1}^M\mathbf{Q}_m\in \mathbb{C}^{N \times K}.\vspace{-0.1cm}
\end{equation}
Problem~\eqref{problem:CRB_X} can be reformulated as \vspace{-0.1cm}
\begin{subequations}\label{problem:CRB_X_Q}
    \begin{align}        
        \min_{ \mathbf{X},\mathbf{Q},\left\{\mathbf{Q}_m\right\}_{m=1}^M}    \!\!&  \text{tr}(\text{CRB}) \\[-0.05cm]
        \mathrm{s.t.}    & \eqref{constraint:PASS}, \\[-0.05cm]
        \label{QH}& \mathbf{Q}_m=\mathbf{H}_m, \forall m \in \mathcal{M}, \mathbf{Q}=\sum\nolimits_{m=1}^M\mathbf{Q}_m,\\[-0.05cm] 
        \label{SINR_Q}& \frac{|\mathbf{q}_{k}^\mathrm{H}\mathbf{w}_{k}|^2}{\sum\nolimits_{i\neq k}|\mathbf{q}_{k}^\mathrm{H}\mathbf{w}_{i}|^2+\mathbf{q}_{k}^\mathrm{H}\mathbf{R}_{s}\mathbf{q}_{k}+\sigma_0^2}\ge\gamma_k.
    \end{align}\vspace{-0.2cm}
\end{subequations}

\noindent The penalty method is adopted to tackle constraints~\eqref{QH}. The resulting  problem can be given as \vspace{-0.1cm}
\begin{subequations}\label{problem:CRB_X_Q_penalty}
    \begin{align}        
        \min_{ \mathbf{X},\mathbf{Q},\left\{\mathbf{Q}_m\right\}_{m=1}^M} \quad \!\!&  \text{tr}(\text{CRB})+\frac{1}{\rho}\Big\|\mathbf{Q}-\sum\nolimits_{m=1}^M\mathbf{Q}_m\Big\|_F^2 \\[-0.05cm]\nonumber&+\frac{1}{\rho} \sum\nolimits_{m=1}^M\left\|\mathbf{Q}_m-\mathbf{H}_m\right\|_F^2\\[-0.05cm]
        \mathrm{s.t.} \quad & \eqref{constraint:PASS},\eqref{SINR_Q} ,
    \end{align}\vspace{-0.6cm}
\end{subequations}

\noindent where $\rho$ is the penalty factor. Problem~\eqref{problem:CRB_X_Q_penalty} is still non-convex due to the non-convex CRB term in the objective function and the non-convex fractional constraint~\eqref{SINR_Q}. The optimization variable matrix $\mathbf{X}$ is highly coupled with the auxiliary matrices $\mathbf{Q}$ and $\mathbf{Q}_m$, the AO technique is adopted to iteratively solve problem~\eqref{problem:CRB_X_Q_penalty}.

\subsubsection{Solving the sub-problem with respect to $\mathbf{Q}$} With fixed $\mathbf{X}$ and $\mathbf{Q}_m, \forall m$, the sub-problem with respect to $\mathbf{Q}$ is\vspace{-0.1cm}
\begin{subequations}\label{problem:CRB_X_Q_penalty_Q}
    \begin{align}        
        \min_{ \mathbf{Q}} \quad &  \Big\|\mathbf{Q}-\sum\nolimits_{m=1}^M\mathbf{Q}_m\Big\|_F^2 \\[-0.05cm]
        \mathrm{s.t.} \quad & \eqref{SINR_Q}.
    \end{align}\vspace{-0.6cm}
\end{subequations}

\noindent Problem~\eqref{problem:CRB_X_Q_penalty_Q} is non-convex due to the non-convex constraint~\eqref{SINR_Q}, which can reformulated as \vspace{-0.1cm}
\begin{equation}
     \sum\nolimits_{i\neq k}\mathbf{q}_{k}^\mathrm{H}\mathbf{W}_{i}\mathbf{q}_{k}+\mathbf{q}_{k}^\mathrm{H}\mathbf{R}_{s}\mathbf{q}_{k}-\frac{1}{\gamma_k}\mathbf{q}_{k}^\mathrm{H}\mathbf{W}_{k}\mathbf{q}_{k}+\sigma_0^2 \le 0.\vspace{-0.1cm}
\end{equation}
To address the non-convexity term, $-\frac{1}{\gamma_k}\mathbf{q}_{k}^\mathrm{H}\mathbf{W}_{k}\mathbf{q}_{k}$, we adopt the SCA method, which iteratively approximates the concave part of the constraint with its first-order Taylor expansion around a feasible point. In the $t$-th iteration, the non-convex term is linearized using its first-order Taylor expansion around a feasible point $\mathbf{q}_{k}^{(t)}$, which can be expressed as\vspace{-0.1cm}
\begin{equation}
\begin{aligned}
        &-\frac{1}{\gamma_k}\mathbf{q}_{k}^\mathrm{H}\mathbf{W}_{k}\mathbf{q}_{k} \le -\frac{1}{\gamma_k}\left(\mathbf{q}_{k}^{(t)}\right)^\mathrm{H}\mathbf{W}_{k}\mathbf{q}_{k}^{(t)} \\[-0.05cm]
        &-\frac{1}{\gamma_k}\mathbf{q}_{k}^\mathrm{H}\mathbf{W}_{k}\left(\mathbf{q}_{k}-\mathbf{q}_{k}^{(t)}\right)-\frac{1}{\gamma_k}\left(\mathbf{q}_{k}-\mathbf{q}_{k}^{(t)}\right)^\mathrm{H}\mathbf{W}_{k}\mathbf{q}_{k}.
\end{aligned}\vspace{-0.1cm}
\end{equation}
By substituting the above into the original constraint, we obtain the following convex surrogate constraint at iteration $t$ as~\eqref{SINR_Q_SCA} at the top of the next page.
\begin{figure*}[!t]
\normalsize
\begin{equation}\label{SINR_Q_SCA}
\begin{aligned}
     \sum\nolimits_{i\neq k}\mathbf{q}_{k}^\mathrm{H}\mathbf{W}_{i}\mathbf{q}_{k}+\mathbf{q}_{k}^\mathrm{H}\mathbf{R}_{s}\mathbf{q}_{k}-\frac{1}{\gamma_k}\mathbf{q}_{k}^\mathrm{H}\mathbf{W}_{k}\big(\mathbf{q}_{k}-\mathbf{q}_{k}^{(t)}\big)-\frac{1}{\gamma_k}\big(\mathbf{q}_{k}-\mathbf{q}_{k}^{(t)}\big)^\mathrm{H}\mathbf{W}_{k}\mathbf{q}_{k}-\frac{1}{\gamma_k}\big(\mathbf{q}_{k}^{(t)}\big)^\mathrm{H}\mathbf{W}_{k}\mathbf{q}_{k}^{(t)}+\sigma_0^2 \le 0.
\end{aligned}\vspace{-0.2cm}
\end{equation}\vspace{-0.1cm}
\hrulefill \vspace*{0pt}
\end{figure*}
In this way, the originally non-convex problem is transformed into a convex one, which can be expressed as\vspace{-0.1cm}
\begin{subequations}\label{problem:CRB_X_Q_penalty_Q_SCA}
    \begin{align}        
        \min_{ \mathbf{Q}} \quad &  \Big\|\mathbf{Q}-\sum\nolimits_{m=1}^M\mathbf{Q}_m\Big\|_F^2 \\[-0.05cm]
        \mathrm{s.t.} \quad & \eqref{SINR_Q_SCA}.
    \end{align}\vspace{-0.6cm}
\end{subequations}

\noindent The approximated problem in each SCA iteration can be efficiently handled using CVX. 

\subsubsection{Solving the sub-problem with respect to $\mathbf{Q}_m, \forall m$} With fixed $\mathbf{X}$ and $\mathbf{Q}$, the sub-problem with respect to $\mathbf{Q}_m, \forall m$,  is
\begin{subequations}\label{problem:CRB_X_Q_penalty_Qm}
    \begin{align}        
        \min_{ \left\{\mathbf{Q}_m\right\}_{m=1}^M }  &   \Big\|\mathbf{Q}\!-\!\!\sum\nolimits_{m=1}^M\!\!\mathbf{Q}_m\Big\|_F^2\!\!+\!\!\sum\nolimits_{m=1}^M\!\!\left\|\mathbf{Q}_m-\mathbf{H}_m\right\|_F^2.
    \end{align}
\end{subequations}
Problem~\eqref{problem:CRB_X_Q_penalty_Qm} is an unconstrained convex optimization problem. Denote the objective function of problem~\eqref{problem:CRB_X_Q_penalty_Qm} as $f\left(\mathbf{Q}_1,\mathbf{Q}_2,\cdots,\mathbf{Q}_M\right)$. The optimal solution is given by \vspace{-0.1cm}
\begin{equation}
       \frac{\partial f}{\partial {\mathbf{Q}_m}} \!=\!\mathbf{Q}_m+\sum\nolimits_{i=1}^M \mathbf{Q}_i-\mathbf{Q}-\mathbf{H}_m\!=\!\mathbf{0}_{N \times K}, \forall m.\vspace{-0.1cm}
\end{equation}
Based on the fact that $\sum_{m=1}^{M}{\partial f\left(\mathbf{Q}_1,\mathbf{Q}_2,\cdots,\mathbf{Q}_M\right)}/{\partial {\mathbf{Q}_m}}=\mathbf{0}_{N \times K}$, we can solve that \vspace{-0.1cm}
\begin{equation}\label{opt_Qm}
        \mathbf{Q}_m^*=\mathbf{B}+\mathbf{H}_m, \forall m,\vspace{-0.1cm}
\end{equation}
where $\mathbf{B}=\frac{1}{M+1}\left(\mathbf{Q}-\sum_{i=m}^M\mathbf{H}_i\right)$.  
\subsubsection{Solving  the sub-problem with respect to $\mathbf{X}$} With fixed $\mathbf{Q}$ and $\mathbf{Q}_m, \forall m$, the sub-problem with respect to $\mathbf{X}$ is\vspace{-0.1cm}
\begin{subequations}\label{problem:CRB_X_Q_penalty_X}
    \begin{align}        
        \min_{ \mathbf{X} } \quad &  \text{tr}(\text{CRB})+\frac{1}{\rho} \sum\nolimits_{m=1}^M\left\|\mathbf{Q}_m-\mathbf{H}_m\right\|_F^2\\[-0.05cm]
        \mathrm{s.t.} \quad & \eqref{constraint:PASS}.
    \end{align}\vspace{-0.6cm}
\end{subequations}

\noindent An element wise optimization is adopted to iteratively optimize the location of each PA while keeping other PAs fixed. The optimization problem with respect to the location parameters of the $m$-th PA on the $n$-th waveguide of PASS $x_m^n$ can be expressed as\vspace{-0.1cm}
\begin{subequations}\label{problem:CRB_X_Q_penalty_X_mn}
    \begin{align}        
        \min_{ x_m^n } \  &  \text{tr}(\text{CRB})+\frac{1}{\rho} \sum\nolimits_{k=1}^K\!\left|\left[\mathbf{Q}_m\right]_{nk}-\left[\mathbf{H}_m\right]_{nk}\right|^2\\[-0.05cm]
        \mathrm{s.t.} \ 
        & x_m^n \in \mathcal{X}  , \forall m,n.
    \end{align}\vspace{-0.5cm}
\end{subequations}

\noindent Problem~\eqref{problem:CRB_X_Q_penalty_X_mn} is a one dimensional optimization problem which can be solved with one dimensional search. Then, the solution of the one dimensional search is \vspace{-0.1cm}
\begin{equation}
     {x_m^n}^*\!\!=\!\mathop{\arg\min}\limits_{ x_m^n\in \mathcal{X}_m^n }  \text{tr}(\text{CRB})+\frac{1}{\rho} \sum\nolimits_{k=1}^K\!\left|\left[\mathbf{Q}_m\right]_{nk}\!-\!\left[\mathbf{H}_m\right]_{nk}\right|^2\!\!.\vspace{-0.2cm} 
\end{equation}

\section{Simulation Results} \label{sec:results}
\begin{figure} [!h]
\centering\vspace{-0.5cm}
\includegraphics[width=0.33\textwidth]{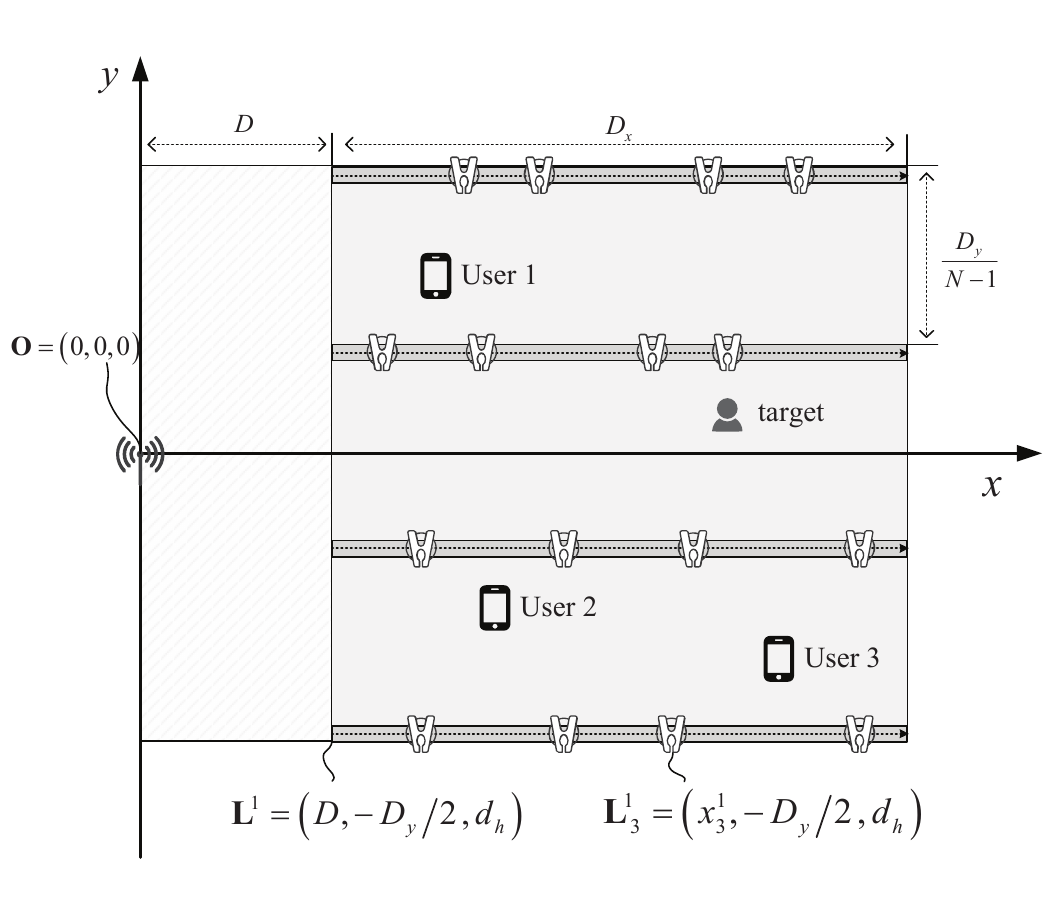}
 \caption{Simulation setup for the PASS assisted ISAC system.
  }\vspace{-0.5cm}
 \label{system_setup}
\end{figure}
\begin{figure*}[!h]
\centering\vspace{-0.3cm}
\subfigure[RCRB versus the power budget of the BS.]{\label{Power_P}
\includegraphics[width=2.15in]{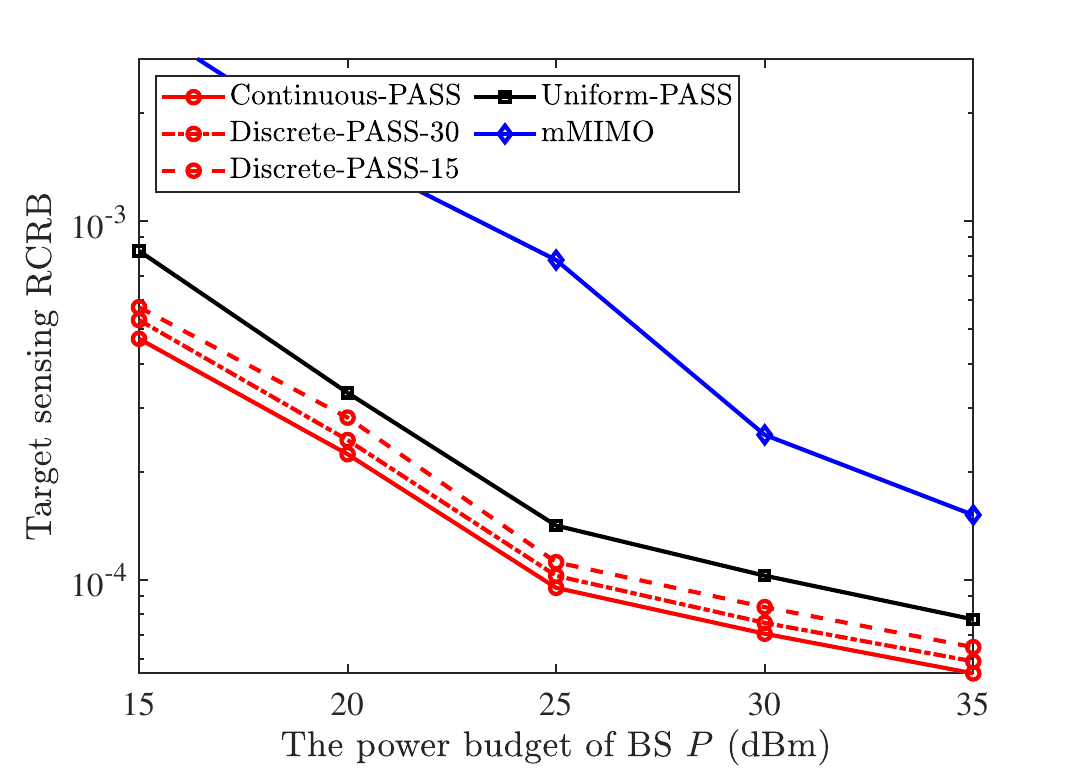}}
\subfigure[RCRB versus the number of waveguide.]{\label{Waveguide_N}
\includegraphics[width=2.15in]{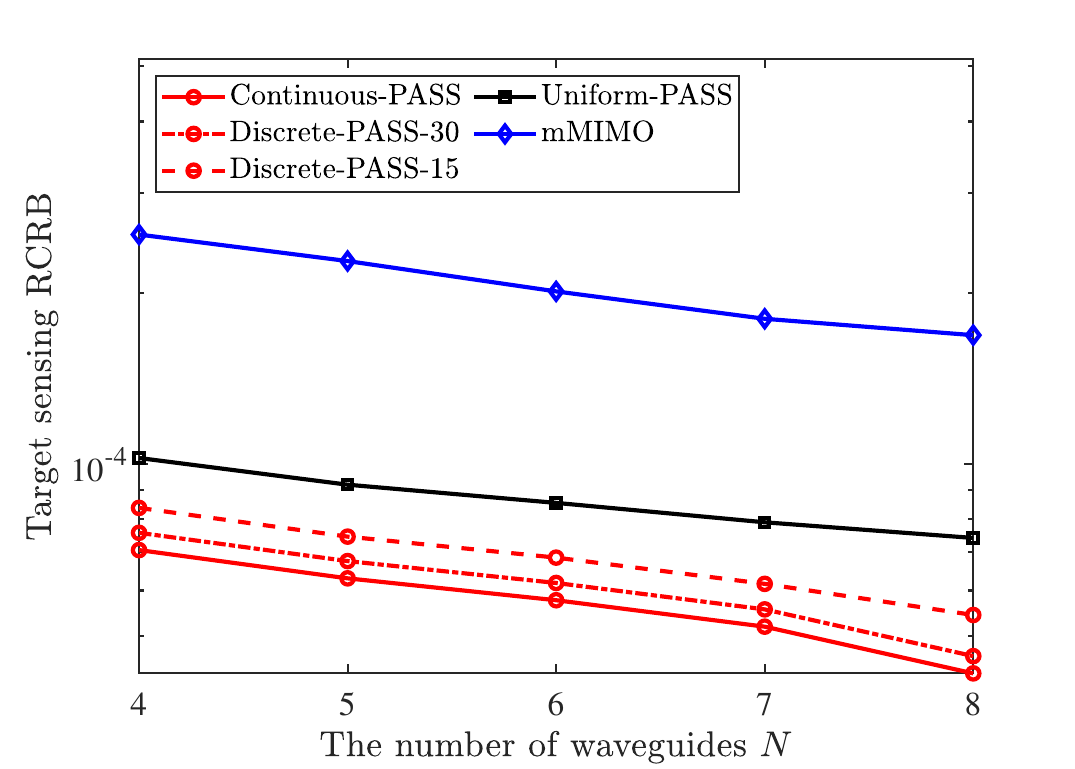}}
\subfigure[RCRB versus the communication SINR.]{\label{SINR_Gamma}
\includegraphics[width=2.15in]{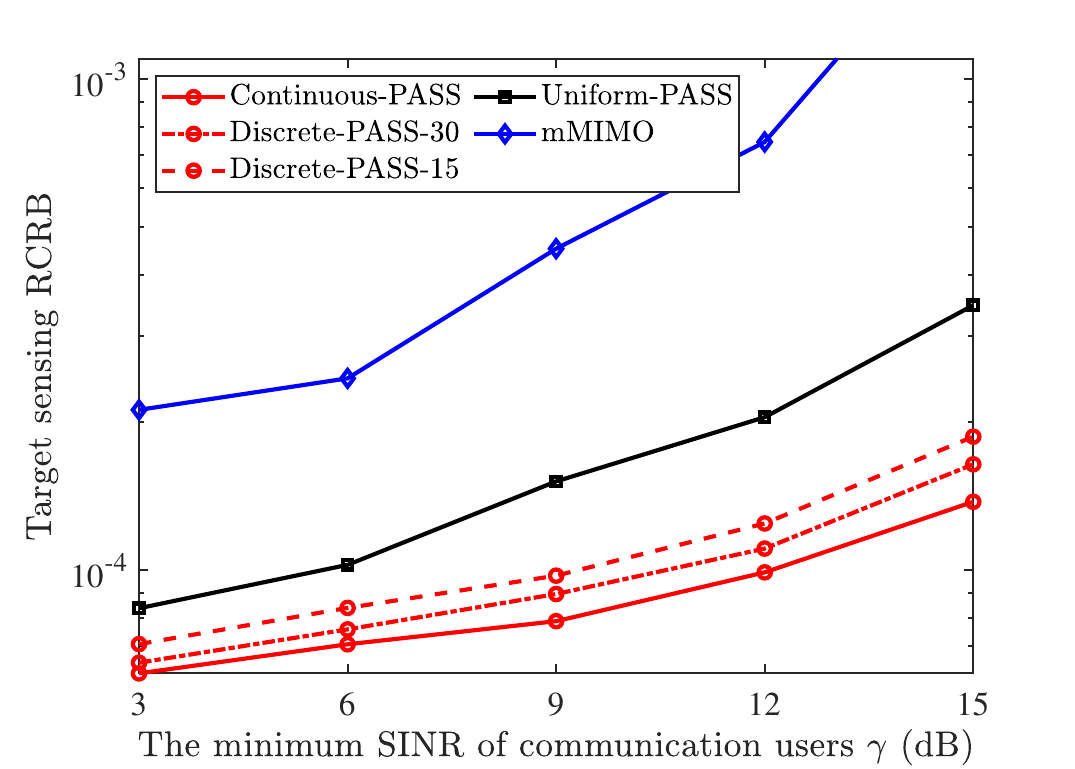}}\vspace{-0.2cm}
\caption{The RCRB of target positioning versus $P$, $N$, $\gamma$.}\label{fig:RCRB_UP}\vspace{-0.4cm}
\end{figure*}
The considered PASS assisted ISAC system is illustrated in Fig.~\ref{system_setup}, where the BS is located at the origin of the Cartesian coordinate system. The service area is defined on the $X$–$Y$ plane, within which all communication users and the sensing target are uniformly distributed. The PASS is equipped with $N=4$ waveguides, each supporting up to $M=4$ PAs, and the minimum separation between adjacent PAs is set to $\delta=\lambda_c/2$. The effective refractive index of each waveguide is $n_e=1.4$, and the attachable length of the waveguides is $L=15$ m. The BS is located at a distance of $D=5$ m from the service area with a PASS height of $d_h=1$ m. The service area spans $D_x \times D_y = 15 \text{m} \times 15 \text{m}$. The BS is equipped with $M_r=8$ receive antennas, and serves $K=3$ single-antenna communication users. The length of the coherent processing interval is $T=256$. The carrier frequency is set as $f_c=28$ GHz. The communication noise power at each user terminal is $\sigma_0^2=-90$ dBm, while the BS receiver noise power is set to $\sigma_s^2=-90$ dBm. The BS has a total transmit power budget of $P=30$ dBm, with a PA power allocation coefficient of $\rho=1/\sqrt{M}$. Each communication user is required to achieve a target SINR of $\gamma_k=\gamma=6$ dB, $\forall k \in \mathcal{K}$. The baseline schemes are introduced as follows:
\begin{enumerate}
\item {\textbf{Massive MIMO ISAC Scheme:} In this baseline scheme,  the BS is equipped with a massive MIMO transmitter. The BS  transmitter adopts  partially connected hybrid beamforming architecture, where the number of radio frequency (RF) chains is set as $N$ while the number of antennas connected to each RF chain is~$M$.}

\item {\textbf{PASS assisted ISAC Scheme with Uniformly Split PAs:} In this baseline scheme, the PAs are uniformly placed along each waveguide. }

\item \textbf{PASS assisted ISAC Scheme with discrete PA design:} In this baseline scheme, the PAs can be deployed to~$Z+1$ predefined discrete positions along each waveguide. 
\end{enumerate}

Fig.~\ref{Power_P} illustrates the variation of the RCRB with respect to the power budget of the BS $P$. As expected, all baseline schemes exhibit improved sensing accuracy when more transmission power is available. The reason is that a higher transmit power leads to a stronger echo signal given the same communication QoS requirement, thereby  tightening the CRB. A comparison among different schemes reveals that PASS assisted ISAC consistently outperforms MIMO-ISAC across the considered power budgets. In particular, the PASS assisted ISAC with continuously deployed PAs achieves the best performance, while the PASS assisted ISAC with discretely deployed PAs with $30$ quantization levels closely approaches it. As the quantization resolution decreases (e.g., to $15$ levels), the performance gap becomes more evident. This degradation stems from the limited pinching beamforming degrees of freedom caused by discrete deployment. Nevertheless, increasing the quantization resolution effectively mitigates this performance loss. The uniform PASS scheme, although inferior to optimized PASS schemes, still significantly outperforms both massive MIMO-ISAC and conventional MIMO-ISAC. This is because, despite its naive beamforming design, the PASS architecture inherently places antennas closer to the communication users and sensing targets.

Fig.~\ref{Waveguide_N} demonstrates the RCRB versus the number of waveguides $N$. It can be observed that, across all considered schemes, the RCRB decreases as the number of waveguides increases, indicating an improvement in sensing accuracy. This improvement is primarily due to the higher number of RF chains available at the BS with more waveguides, which provides greater spatial degrees of freedom. As a result, the system can perform more flexible beamforming for both communication and sensing, thereby enhancing the overall sensing performance.

Fig.~\ref{SINR_Gamma} illustrates the relationship between the RCRB and the minimum communication SINR $\gamma$ of the users. As the minimum communication SINR requirement increases, the system RCRB also increases, indicating a degradation in sensing accuracy.  This phenomenon can be attributed to the finite power and communication QoS constraints at the BS. Higher SINR requirements necessitate allocating more power and resources to the communication tasks to satisfy stringent  QoS  demands. Consequently, fewer resources remain available for sensing, leading to a higher RCRB and thus lower sensing precision.

\section{Conclusions}\label{sec:conclusions}
In this work, a PASS assisted ISAC system was investigated, where a sensing CRB minimization problem was formulated under communication QoS constraints to explore the fundamental sensing accuracy limits. To address the overall non-convex optimization problem, an AO based algorithm was developed, which iteratively solves the two sub-problems using the SDR method. Simulation results validated the effectiveness of the proposed framework, demonstrating its superior performance over benchmark schemes.\vspace{-0.1cm}

\section*{Appendix~A\\Derivation of the FIM for PASS assisted ISAC} \label{Appendix:A}
\renewcommand{\theequation}{A.\arabic{equation}}
\setcounter{equation}{0}
Reformulate the received PASS signal over a coherent processing interval including $T$ time slots as vectorized signal
\begin{equation}
         \text{vec}\left(\mathbf{Y}_s\right)=\text{vec}\left(\mathbf{G}\mathbf{S}\right)+\text{vec}\left(\mathbf{N}_s\right)\sim\mathcal{CN}\left(\mathbf{u}, \mathbf{R}_n\right),
 \end{equation} 
where $\mathbf{u}=\text{vec}\left(\mathbf{G}\mathbf{S}\right)$, $\mathbf{R}_n=\sigma_s^2\mathbf{I}_{M_rT}$. The element on the $i$-th row and the $j$-th column of $\mathbf{J}$ can be expressed as
\begin{equation}\label{FIM_calculate}
\begin{aligned}
        \left[\mathbf{J}\right]_{ij}=\frac{2}{\sigma_s^2}\Re\left(\frac{\partial \mathbf{u}^\mathrm{H}}{\partial \eta_i} \frac{\partial \mathbf{u}}{\partial \eta_j}\right), \forall i,j \in \{1,2,3,4\},
\end{aligned}
\end{equation}
where $\eta_i$ is the $i$-th element of $\boldsymbol{\eta}$. By defining $\mathbf{A}=\mathbf{a}\left(\theta\right)\mathbf{h}_t^\mathrm{H}$, it follows that

\begin{equation}
\begin{aligned}
        \!\!\!\frac{\partial \mathbf{u}}{\partial \boldsymbol{\eta}}\!\!=\!\!\left[\beta\text{vec}(\dot{\mathbf{A}}_{x_t}\mathbf{S}),\beta\text{vec}(\dot{\mathbf{A}}_{y_t}\mathbf{S}),\text{vec}(\mathbf{A}\mathbf{S}),j\text{vec}(\mathbf{A}\mathbf{S})\right]\!\!,
\end{aligned}
\end{equation}
where $\dot{\mathbf{A}}_{x_t}={\partial \mathbf{A}}/{\partial x_t}$ and $\dot{\mathbf{A}}_{y_t}={\partial \mathbf{A}}/{\partial y_t}$.
Matrix $\mathbf{J}_{11}$ is
\begin{equation}\label{eqn:FIM_J11}
        \mathbf{J}_{11}=\begin{bmatrix}
         {J}_{x_tx_t} & {J}_{x_ty_t}; \quad
         {J}_{x_ty_t} & {J}_{y_ty_t} 
    \end{bmatrix}.
\end{equation}
Based on~\eqref{FIM_calculate}, the element ${J}_{i_tj_t}$, $\forall i,j\in\left\{x,y\right\}$, is
\begin{equation}\label{eqn:FIM_J11_ele}
      {J}_{i_tj_t}=\frac{2\left|\beta\right|^2T}{\sigma_s^2}\Re\left(\text{tr}\left(\dot{\mathbf{A}}_{j_t}\mathbf{R}\dot{\mathbf{A}}_{i_t}^\mathrm{H}\right)\right),
\end{equation} 
The matrix $\mathbf{J}_{12}$ can be expressed as 
\begin{equation}\label{eqn:FIM_J12}
\begin{aligned}
        \mathbf{J}_{12}\!=\! \frac{2 T}{\sigma_s^2} \Re \!\Big(\beta\!\left[ \mathrm{tr}(  \dot{\mathbf{A}}_{x_t} \mathbf{R}   {\mathbf{A}}^\mathrm{H} ),  \mathrm{tr}(  \dot{\mathbf{A}}_{y_t} \mathbf{R}  {\mathbf{A}}^\mathrm{H} )\right]^\mathrm{H} \!\!\left[1, j\right]\Big) .
\end{aligned}
\end{equation}
The matrix $\mathbf{J}_{22}$ can be expressed as 
\begin{equation}\label{eqn:FIM_J22}
        \mathbf{J}_{22}=\frac{2T}{\sigma_s^2}\mathbf{I}_{2}\Re\left(\text{tr}\left({\mathbf{A}}\mathbf{R}{\mathbf{A}}^\mathrm{H}\right)\right).
\end{equation}

\bibliographystyle{IEEEtran}
\bibliography{myref}

\end{document}